\documentclass{elsart}
\usepackage{epsfig}

\begin{document}
\newcommand*\mycite[1]{\cite{#1}}
\newcommand*\myref[1]{(\ref{#1})}
\newcommand*\mylabel[1]{\label{#1}}

\begin{frontmatter}
\title{Application of Zhangs Square Root Law and 
       Herding to Financial Markets}
\author{Friedrich Wagner}
\address{Institut f\"ur Theoretische Physik,
Universit\"at Kiel, Leibnizstrasse 15, D-24098 Kiel, Germany, e-mail:
wagner@theo-physik.uni-kiel.de }
\begin{abstract}
We apply an asymmetric version of  Kirman's herding model to
volatile financial markets. In the relation between returns and
agent concentration we use the square root law proposed by Zhang.
This can be derived by extending the idea of a critical mean field
theory suggested by Plerou et al.
We show that this model is equivalent to the so called 3/2-model of
stochastic volatility. The description of the unconditional distribution
for the absolute returns is in good agreement with the DAX independent
whether one uses the  square root or a conventional linear relation. Only the
statistic of extreme events prefers the former. The description of 
the autocorrelations are in much better agreement for the square root law.
The volatility clusters are described by
a scaling law for the distribution of returns conditional
to the value at the previous day in good agreement with the data.
\end{abstract}

\end{frontmatter}

\section{Introduction} \label{intro}
Any model for the price or index of a financial market should
account for the following so called stylized facts \mycite{c1,c2}.
(i): The sign of the return cannot be predicted. In most models
   this fact is build in by assuming the return proportional to an iid
   noise.
(ii):  The probability for having an absolute return larger than $x$ decays for
   large $x$ as a power law $x^{-n_T}$. The tail index $n_T$ seems to be
   universal \mycite{uni} in the range 3-4. A precise determination of $n_T$ for
   dayly returns is hampered 
   by the impossibility to reach the asymtotic regime in the data. Therefore
   one should require a good description of all returns.
(iii):  Returns are not independent. They form so called
   volatility cluster. A quantitative measure of this effect may be the
   distribution of first passage times \mycite{c3}. In this paper we
   use the distribution of the returns conditional to the value of the
   previous day.
(iv):  Absolute returns are correlated. In the case of high frequency
   data a hyperbolic decay of the autocorrelation 
   indicates a long memory process. For dayly returns it behaves
   neither as a power law nor a simple exponential expected from
   a Markov process.\\
To acchieve these properties one can distinguish two schools. The first are
the stochastic volatility models. Here one assumes the return to be an iid
noise multiplied with
a volatility factor $\sqrt{V_T}$, where $V_T$ typically
follows a Markov process. Examples
are the GARCH model \mycite{c5} and its continuous time version \mycite{c6}.
Other posibilities are the CIR-model \mycite{CIR} or
the 3/2 model \mycite{Ahn} discussed in ref \mycite{c9}. The multifractal
model \mycite{c10} implements fact (iv) by occurence of many   time
scales. Any model for dayly returns faces the following
difficulty. The quality of
data for dayly returns does not allow to estimate reliably more than three 
parameters, for example a scale parameter given by the first moment,
a combination of parameters describing the tail index and a rate constant
for the time dependence as in GARCH(1,1). Therefore the parsimonity
principle invoked by Calvet and Fisher \mycite{c10} is rather 
a must and not a freedom.
For stochastic volatility models a second difficulty arrises, that
the choice of the free parameters is
dictated by mathematical convenience. Their relation to market properties
remains obscure. This difficulty is avoided in agent or microscopic
models (for a recent review see ref \mycite{c11} and the references
given therein). The behaviour of
the agents may be characterized by utility functions \mycite{c12}. The
observed power law (fact (ii)) motivates the use of critical physical
models as the application of percolation by Cont-Bouchaud \mycite{CB}.
The percolation cluster are interpreted as  herding behaviour of the agents.
In the Lux-Marchesi model \mycite{LM} a herding mechanism appears 
explicitly in the transition probabilities
between chartists and fundamentalistic traders. This leads to an intermittent
behaviour of agent numbers and thereby also for the returns.   
To see whether herding alone can describe the stylized facts a symmetric
Kirman model \mycite{Kir} has been studied \mycite{AL}. Quantitative
agreement with the data could be acchieved \mycite{AWL} within an asymmetric
Kirman model. In the  latter agents differ slightly from those in 
the Lux-Marchesi model that the two types of chartists (optimists
and pessimists) are replaced by one sort of agents (noise traders). The
difference between optimists and pessimists is described by an iid 
random number. With this approximation the return behaves similar to a
stochastic volatility model, except $\sqrt{V_T}$ follows a Markov process
instead of $V_T$. In this paper we apply the Kirman model with a further
simplification. In empirical investigations it has been found
\mycite{AWL,Alf} that fundamentalists are much less affected by herding
than noise traders, which implicates a very asymmetric herding model. In
the limit of large asymmetry the number of free parameters is reduced
from three to two. In the spirit of parsimonity we adopt this approach.\\
For an application to financial markets an equation of state is needed
relating the return to ratios of agent numbers. In a Walrasian market
\mycite{AWL} the volatility factor $\sqrt{V_T}$ depends linearely on
the concentration of agents. The model using the simplified Kirman
model and this linear relation has been studied in \mycite{Alf}. In this
paper we adopt another equation of state. In contrast to a Walrasian market
the time for an agreement of the agents on a common return ought to be
finite. Assuming a random walk Zhang \mycite{Zhang,Stauf}
concluded that $V_t$ itself should be proportional to the agent concentration. 
We use this approach in the present paper. It
offers two desirable properties. Since in stochastic volatility models
$V_t$ and in herding models the concentration proportional to
$V_t$ follow simple Markov processes,
one might be able to show the equivalence of the latter with one of the
stochastic volatility models. Secondly one can use the empirical observation of
Plerou et al \mycite{Stan3}, that the return and the imbalanced volume
satisfy the same relation as the order parameter and an external field
in a critical mean field theory. In the framework of a universal $\Phi^4$
\mycite{Crit} there ought to be another equation for the return at zero
field. We will show that Zhang's law emerges within such a model
naturally from risk aversity.
Whether the Walrasian linear relation or Zhang's law are correct can be decided
at present only empirically. We will compare our model and the results from ref
\mycite{Alf} with the probability density function (hereafter abbreviated
with pdf ) and the autocorrelation function with the empirical returns
derived from the DAX.\\
The paper is organized in the following way. In section \ref{sec1} we describe
our version of the Kirman model and in section \ref{sec1a} a derivation
of Zhang's law. The
resulting unconditional pdf and the common pdf for two absolute returns
at different times
are given in section \ref{Prob}. The square root law together with a 
Gaussian noise allows analytical calculation of various observables. 
As examples we treat the autocorrelation function and the pdf conditional on the value of the previous day. These predictions we confront with the data in
section \ref{Data}. In section \ref{concl} we make some concluding remarks.

\section {Kirman's Herding Model \label{sec1} }
We want to describe the changes of prices in
 a stock market with different agents
which can change their strategy as function of time. Apart from the
specification of an agent model a relation between the relative changes
(return) of the stock index and the agent numbers is needed which will
be treated in the next section. As herding model
we use the Kirman model \mycite{Kir} 
discussed by Alfarano et al \mycite{AWL}. It is
defined by the transition probability between two types of behaviour.
$S$ denotes the mood of noise traders and $F$ a conservative or fundamentalistic
strategy. The probabilities per unit time to change opinion are given by
\begin{eqnarray} \mylabel{e111}
\pi(S\to F)&=&(a_2+b\cdot n_F) \nonumber \\
\pi(F\to S)&=&(a_1+b\cdot n_S) \quad .
\end{eqnarray}
The parameters 
$a_{1,2}$ correspond to spontaneous changes of mind. The second terms in 
equ \myref{e111}  are proportional to the agent numbers $n_{S,F}$
in the new state and therefore
describe the herding effect. Despite of the extensive
character of $\pi$ this Markov process leads for a large number of agents
to a non trivial equilibrium distribution for intensive ratios as 
$u \propto n_F/(n_S+n_F)$. As shown in \mycite{AWL}
the master equation belonging to \myref{e111} leads to a
Fokker Planck equation (hereafter  abreviated by FPE) for the
 pdf $w(u,t)$ and a Langevin equation
for $u(t)$. In application of the model it turned out \mycite{AWL,Alf}
 that in most
cases $a_1>>a_2$ holds. This means that fundmentalists are much less
influenced by herding than noise traders. Invoking the parsimonity 
principle we consider the limit of large $a_1/b$. For the ratio
\begin{equation} \mylabel{e113a}
u=\frac{a_1}{b}\cdot \frac{n_F}{n_F+n_S}
\end{equation}
it is shown in appendix \ref{A.1} that one obtains the following FPE
\begin{equation} \mylabel{e113}
\frac{\partial w(u,t)}{\partial t}=b\; \frac{\partial}{\partial u}
       \left [u-\beta -1+\frac{\partial}{\partial u}u\right ]w(u,t) \quad ,
\end{equation}
with the parameter $\beta=a_2/b-1$ describing the equilibrium pdf. 
This implies, that the number of fundamentalists is much smaller than
that of the noise traders.  If we transform the
Langevin equation belonging to \myref{e113} to the variable $y=1/u$
\begin{equation} \mylabel{e114}
\Delta y=b\;\Delta t \;y_t\;\left (1+(1-\beta)y_t\right )\;
          +\; (2b \Delta t \;y_t^3)^{1/2}\eta_t \quad ,
\end{equation}
we obtain a well known model of stochastic volatility, the so called 3/2 model
\mycite{Ahn} provided we identify $y_t$ 
in the relation between the return $r_t$ and a Gaussian noise $\eta_t$
\begin{equation} \mylabel{e115}
r_t=\sqrt{V_t}\cdot \eta_t
\end{equation}
with the volatility factor $V_t$. Equ \myref{e114} corresponds to a
special version of the 3/2 model where drift and diffusion constants are
related to ensure existence of an equilibrium pdf for $y$.
From $V_t=u_t$ follows 
another frequently used model \mycite{CIR}.\\
The limit of large $a_1/b$ affects also the time scales. As shown in the 
appendix the process \myref{e111} should be applied to micro time steps
$\Delta t_m$, whereas the variable $u$ changes with the observed  time scale
$\Delta t$. Both are related by $\Delta t=a_1 \Delta t_m /b$.
Eliminating $a_1/b$ with \myref{e113a} we get
\begin{equation} \mylabel{e115a}
\frac{n_F}{n_F+n_S}=u\cdot \frac{\Delta t_m}{\Delta t} \quad ,
\end{equation}
which shows an explicit dependency of the agent ratio 
on the time resolution $\Delta t$.\\
As shown in \mycite{Ahn} the FPE can be solved analytically for
the equilibrium pdf $w_0(u)$
\begin{equation} \mylabel{e116}
w_0(u)=\frac{u^\beta}{\Gamma(\beta+1)}\exp(-u)
\end{equation}
and for the conditional pdf $w(u,t+t_0|\bar{u},t_0)=A(u|\bar{u},\exp(-bt))$ with
\begin{equation} \mylabel{e117}
A(u|\bar{u},z)=w_0(u)\; \frac{(u\bar{u}z)^{-\beta/2}}{1-z} \;
               \exp\left (-\frac{z(u+\bar{u})}{1-z}\right ) \; 
               I_\beta\left (\frac{2\sqrt{u\bar{u}z}}{1-z}\right )\quad .
\end{equation}
$I_\beta$ denotes the Bessel function with imaginary argument. The
kernel $A$ has the following convolution property
\begin{equation} \mylabel{e118}
\int d\bar{u}A(u|\bar{u},z)A(\bar{u}|u',z')=A(u|u',z\cdot z')
\end{equation}
and can be used for the time evolution of $w(u,t+t_0)$
\begin{equation} \mylabel{e119}
w(u,t+t_0)=\int d\bar{u}A(u|\bar{u},\exp(-bt))w(\bar{u},t_0)\quad .
\end{equation}

\section{Derivation of Zhang's Law  \label{sec1a} }
The problem of this section deals with relation of agent ratio to
the observable return $r_t$. The small decay time in the autocorrelation 
of the returns
suggests that the dynamic of the price is much faster than the change of 
agent numbers. The extreme case of an instantaneous Walrasian market has been
assumed in ref \mycite{AWL} to model the behaviour of agents in the
Lux-Marchesi model\mycite{LM}. With the additional assumption that the change
between optimistic and pessimistic noise traders occurs also on a
fast time scale one gets \mycite{AWL} a linear relation between the return
and the agent ratio $n_S/n_F$. Using $u$ for small values of $n_F$ one gets
\begin{equation} \mylabel{e122}
r_t=\Delta \ln S_t=r_0\cdot \frac{1}{{u_t}}\cdot \eta_t
\end{equation}
The noise $\eta_t$ incorporates many 
changes during $\Delta t$ due to the noise traders.\\
In this Walrasian approach the price change is set instantaneously, a
closed order book is assumed and the agents have to accept any new return
irrespectively of their risk aversity. Considering for example the XETRA market
these assumptions seem to be questionable. Removing the first assumption
Zhang \mycite{Zhang} assumed
the time needed to get the new price is proportional to the demand, which
is linear in the agent numbers. Describing the evolution of the price
by a random walk the return will be proportional to the square root
of the agent numbers. Therefore the linear relation \myref{e122} 
should be replaced by the so called sqare root law given by
\begin{equation} \mylabel{e121}
r_t=r_0\cdot \frac{1}{\sqrt{u_t}}\cdot \eta_t
\end{equation}
It has been applied \mycite{Stauf} in context of the percolation model
of Cont-Bouchaud \mycite{CB} in order to obtain a tail index larger than 2.
Equ \myref{e121} agrees with the return obtained in the 3/2 stochastic model.
Therefore the parameters of the latter can be related to the
behaviour of agents.\\
To obtain a less qualitative derivation of \myref{e121} and to include
the effect of risk aversity and an unbalanced order book we start with an
observation made by Plerou et al \mycite{Stan3}, who showed empirically
that the absolute return $v$ increases with the imbalanced volume $\Omega$ 
at small time intervalls $\Delta t$ and small $\Omega$ as
\begin{equation} \mylabel{e123}
v_t=|r_t|\propto \Omega_t^{1/\delta}
\end{equation}
with an exponent $\delta =2.9\pm 0.3$. At large $\Delta t$ and $\Omega$
the return saturates as $v=B_0\tanh (B_1\Omega) $. As suggested already 
in ref \mycite{Stan3}
this could be interpreted as an equation of state in a critical mean
field theory if $v$ corresponds to a scalar order parameter and $\Omega$
to an external field. As dynamical fields in such a theory we use the
return $\Phi_{k,i}$ expected by agent $i$ of type $k=F,S$. After time
$\Delta t$ equilibrium is reached at a common return $r=E[\Phi_{k,i}]$.
A critical mean field theory with a scalar order parameter is described
by the universal $\Phi^4$ model \mycite{Crit}. This is defined by
the following cost or energy functional 
\begin{equation} \mylabel{e123a}
S(\Phi) = S_{\mbox{int}}(\Delta \Phi)\; + \;
        \sum_{k=S,F}\sum_{i} \left [\frac{\beta_k}{4} \; \Phi^4_{k,i}
        -  \frac{\alpha_k}{2}\;\Phi^2_{k,i}
         -\Omega \Phi_{k,i} \right ] \quad .
\end{equation}
The first term $S_{\mbox{int}}$ is a positive definite quadratic form in the 
differences $\Delta \Phi_{ik,jl}=\Phi_{k,i}-\Phi_{j,l}$  
describing the interaction between the agents.
We assume it strong enough that in equilibrium the mean field value for
$\Phi_{i,k}$ is independent of $i,k$. This means the agents reach a common value
$v$ for their expected absolute return.  We normalize $S_{\mbox{int}}$ by 
$S_{\mbox{int}}(0)=0$. The  $\Phi^4$ terms lead to large cost functions for
large $\Phi$. They account for the risk aversity of the agents. 
We expect that noise traders are much less affected by risk aversity than
fundamentalists and can set $\beta_S$ to zero. Since the disordered phase
$E[\Phi]=0$ is not observed, the coefficients $\alpha_k$
in the quadratic term in 
\myref{e123a} must be positive. A negative contribution to the cost function
of the noise traders can be interpreted as 
'happy loser' effect \mycite{Zhang}, that fundamentalists are willing to accept
a loss of money restricted by their risk aversity. The analogous term for the 
fundamentalists can be neglected by setting $\alpha_F=0$. The last term 
in \myref{e123a} corresponds to the coupling of the return
to an imbalanced volume.
If this coupling is independent of $k$ the coefficient can be set without
loss of generality to 1. In the limit of large agent numbers
the model is solved by a constant $\Phi_{k,i}=v$ with negligeable variance. 
The value of $v$ is given by the minimum of \myref{e123a}.
With the above discussed approximations $S$ reads
\begin{equation} \mylabel{e123b}
S(v)=\frac{\beta_F}{4}\;n_F\; v^4 - \frac{\alpha_S}{2}n_S\; v^2
     -\Omega v (n_S+n_F)\quad .
\end{equation}
Introducing the parameter $\Omega_0$ by
\begin{equation} \mylabel{e123b1}
\Omega_0=\left [\frac{(n_S\alpha_S)^3}{(n_S+n_F)^2\beta_F n_F} \right ]^{1/2}
\end{equation}
the value of $v$ making $S(v)$ to a minimum satifies the following
mean field equation
\begin{equation} \mylabel{e123b11}
\left (\frac{n_S\alpha_S}{(n_S+n_F)\Omega_0}\; v\right )^3\; -\; 
     \frac{n_S\alpha_S}{(n_S+n_F)\Omega_0}\; v\; -\; \frac{\Omega}{\Omega_0}=0
\end{equation}
The solution of \myref{e123b11} can be written as
\begin{equation} \mylabel{e123b2}
\frac{v}{\Omega_0}=\frac{n_S+n_F}{\alpha_S n_S}\cdot 
   f\left (\frac{\Omega}{\Omega_0}\right )
\end{equation}
Equ  \myref{e123b2} expresses the scaling property of the critical theory.
$\Omega$, $\Omega_0$ and $v$ may be rescaled by the same factor without
altering the results. The scaling function $f$ does not depend on any 
parameter. Its behaviour near $x\sim 0$ and $x\to \infty$ is given by
\begin{equation} \mylabel{e123b3}
f(x)=\left\{ \begin{array}{l@{\mbox{\quad for\quad }}l}
            1+\frac{1}{2}x +\cdots & x \ll 1 \\
            (x)^{1/3}(1+(x)^{-2/3}+\cdots) & x \gg 1 \end{array} \right .
\end{equation}
If only the leading terms are kept, for historical reasons the first form
is called the zero field equation of state and the second the critical
isotherm. The latter written in $v$ reads as
\begin{equation} \mylabel{e123c}
v^3=\frac{n_S+n_F}{\beta_F\; n_F}\cdot \Omega\quad ,
\end{equation}
where we recover \myref{e123} for $\Omega \gg \Omega_0$. For 
$\Omega \ll \Omega_0 $ we get another equation of state
\begin{equation} \mylabel{e123d}
v=\sqrt{\frac{\alpha_S}{\beta_F}\cdot \frac{n_S}{n_F}}\quad ,
\end{equation}
which reproduces the square root law of Zhang. The size of $\Omega_0$
depends on the time resolution $\Delta t$, if we use the Kirman model of
the previous section. Inserting equ \myref{e115a} for the agent ratios we get
\begin{equation} \mylabel{e123d1}
\Omega_0\propto \left (\frac{\Delta t}{u}\right )^{1/2} \quad .
\end{equation}
Therefore the empirical relation \myref{e123} should hold only at small time
scales in agreement with the data \mycite{Stan3}. For large time scales
the Zhang's law ought to be applied. In addition we have the prediction, that
the rate factor $r_0$ in \myref{e121} should vary with $\Delta t$ as
\begin{equation} \mylabel{e123d2}
r_0\propto \left (\Delta t\right )^{1/2} \quad .
\end{equation}
This power law is in nice agreement with the observed \mycite{Stan1} power of
0.51.\\
There are still two problems within our derivation of the square root law.
The application of a critical theory is confined to small values of the
fields. At finite values the scaling law \myref{e123b2} may no longer
valid. Secondly the return does not saturate as required by the data,
but will increase as \myref{e123c} for large $\Omega$. There exist of course
many models which have the same small field expansion as \myref{e123a}.
Even if we restrict ourselves to a minimal model of Ising spins which
has been used in ref \mycite{Stan3} to describe the saturation there will
be an extra parameter for the distance from the scaling law \myref{e123b2}.
As shown in appendix \ref{spin} for a spin model the mean field equation
\myref{e123b11} should be replaced by
\begin{equation} \mylabel{e123d3}
\frac{n_S\alpha_S\; v}{(n_S+n_F)\Omega_0}=\;\frac{3}{(3-y^2)y}\cdot
   \tanh\left (\frac{n_S\alpha_S\; vy}{(n_S+n_F)\Omega_0} \;+\;
   \frac{y^3}{3}\cdot \frac{\Omega}{\Omega_0} \right )
\end{equation}
with a parameter $y$ describing the deviation from the critical region. For
$y\to 0$ we recover the critical mean field equation. Finite 
$y$ may be adjusted to the observed saturation of the return with the 
unbalanced volume $\Omega$. Therefore the critical isotherm \myref{e123c}
is very model dependent. On the other side the zero field equation for
$\Omega \ll \Omega_0$ is almost unchanged. In the case of \myref{e123d3} we 
obtain
\begin{equation} \mylabel{e123d4}
v=a(y)\cdot\sqrt{\frac{\alpha_S}{\beta_F}\cdot \frac{n_S}{n_F}}
\end{equation}
with a $y$-dependent constant $a(y)$ which is the non zero solution of
\begin{equation} \mylabel{e123d5}
a\cdot y=\frac{3}{3-y^2}\cdot \tanh(a\cdot y) \quad .
\end{equation}
Since this constant can be absorbed into the rate constant $r_0$,
the square root
law appears to be robust against deviation from criticality. In appendix 
\ref{spin} we also replaced the Ising spin model by a compact model, 
which leads to the same conclusions.\\
At present stage of research
only empirical observations can decide on the possibilities \myref{e121}
or \myref{e122}. In the present paper
we want to discuss whether there are advantages of the square root 
law \myref{e121} over the linear relation \myref{e122}.

\section{Unconditional probability and 2-point function \label{Prob} }
Suppose one observes a time serie of
the absolute returns $H_t=[v_1,\cdots,v_t]$. We assume the time $t$ to be
an integer by absorbing the time step $\Delta t$ in the rate factor $b$.
Only the combined process
$[H_t,U_t]$ with the history $U_t=[u_0,\cdots,u_t]$ of the herding
variables is a Markov process. This property is expressed by the following
recursion formula for the pdf $f(H,U)$
\begin{equation} \mylabel{e201}
f(H_t,U_t)=g(v_t|u_t)\cdot A(u_t|u_{t-1},e^{-b})\cdot f(H_{t-1},U_{t-1})\quad .
\end{equation}
Assuming the sqare root law \myref{e121} with a Gaussian noise the 
conditional distribution $g$ is given by
\begin{equation} \mylabel{e202}
g(v|u)=N_0\sqrt{u}\cdot \exp(-\frac{v^2u}{2r_0^2})
\end{equation}
with the normalization constant $N_0=\sqrt{2/\pi r_0^2}$.
To calculate the pdf for the absolute returns $H_t$ the solution of the
recursion \myref{e201} has to be integrated over all $u$ variables.
 The sqare root law 
leads to an exponential dependence of $g$ as function of $u$, which matches 
with the exponentials in $A$ allowing analytical calculations. In principal
the recursion \myref{e201} can be used to compute $f(H_t)$. The singularity
of $A$ for $\exp(-b)\sim 1$ presents a problem, if one encounters a small value
of the rate $b$. For the determination of the
parameters $\beta,r_0$ and $b$ two simpler cases are sufficient and we
defer an approximative solution of the recursion \myref{e201} to a future
publication \mycite{AW2}.\\
The common pdf $f(v_{t+t_0},v_{t})$ for the absolute returns at $t_0$ and
$t+t_0$ is obtained from the $t$ times iterated recursion 
\myref{e201} by integrating over $U_t$ and all
intermediate $v_{t+1}\cdots v_{t-1+t_0}$. The latter integration 
eliminates the corresponding
$g$-factors and we can use the convolution property \myref{e118} of $A$ to
perform the intermediate $u$ integration leading to
\begin{eqnarray} \mylabel{e203}
f(v_{t+t_0},v_{t_0})&=&\int du\;du_{t_0}g(v_{t+t_0}|u)A(u|u_{t_0},e^{-bt})
                      \nonumber \\
                    &\;\; &\int dH_{t_0-1}dU_{t_0-1}f(H_{t_0},U_{t_0})\quad .
\end{eqnarray}
For large $t$ the operator $A$ converges to the equilibrium distribution
$w_0(u)$ independent of $u_{t_0}$. Therefore $f(v_{t+t_0},v_{t_0})$ 
factorizes in the limit of large $t$ as
\begin{equation} \mylabel{e204}
f(v_{t+t_0},v_{t_0})=G_0(v_{t+t_0})\cdot f(v_{t_0})
\end{equation}
with the equilibrium pdf
\begin{equation} \mylabel{e205}
G_0(v)=\int du\;g(v|u)\;w_0(u)=N_0\frac{\Gamma(\beta +3/2)}{\Gamma(\beta +1)}
       \; \left (1+\frac{v^2}{2r_0^2} \right )^{-\beta -3/2}
\end{equation}
This Pareto like distribution becomes at large $v$ 
a power law $v^{-n_T-1}$ with a
tail index $n_T$ determined by the herding parameter
\begin{equation} \mylabel{e206}
n_T=2\beta+2\quad .
\end{equation}
From equ \myref{e205} all equilibrium moments may be calculated ( see
equ \myref{A21} in the appendix). For example
the second moment
\begin{equation} \mylabel{e207}
E_0[v^2]=\frac{r_0^2}{\beta}
\end{equation}
will be used for expressing $r_0$ in terms of $\beta $ and the observable 
second moment in equilibrium
\footnote{Expectation values taken with the equilibrium pdf \myref{e205} will
be denoted by $E_0[\; .\;]$   }.\\
To derive an explicit form of the two point function $f(v_{t+t_0},v_{t_0})$
we take in equ \myref{e203} large $t_0$ with finite $t$. Analogue to
the previous case the distribution factorizes and we obtain
\begin{equation} \mylabel{e208}
f(v_{t+t_0},v_{t_0})=\int\; dudu'g(v_{t+t_0}|u)A(u|u',e^{-bt})\;
                     g(v_{t_0}|u')w_0(u')\quad .
\end{equation}
The integrations in \myref{e208} can be carried out and $f(v_{t+t_0},v_{t_0})$
is expressed in terms of hypergeometric functions. The exact form is given
in appendix \ref{A.2}. An interesting property of $f$ is that it depends
only on a combination $\zeta(v_{t+t_0},v_{t_0})$
but not each of the variables $v_{t_0}$ and $v_{t+t_0}$
separately. Such a behaviour implies scaling laws. 
We demonstrate this scaling law in
the simple case of $t=1$ and neglected terms of order $b $. 
As shown in appendix \ref{A.2}
the conditional pdf $f(v_{t+1}|v_{t})$ is given by
\begin{equation} \mylabel{e209}
f(v_{t+1}|v_{t})=\sqrt{\frac{2}{\pi(2r_0^2+v_t^2)}}\cdot 
             \frac{\Gamma(\beta +2)}{\Gamma(\beta +3/2)}
             \left (1+\frac{v^2_{t+1}}{2r_0^2+v_t^2} \right )^{-\beta -2}\quad .
\end{equation}
If we eliminate $r_0$ with equ \myref{e207} and introduce the scaling
variable $x_{t+1}$ by
\begin{equation} \mylabel{e209a}
x_{t+1}=\frac{v_{t+1}}{\sqrt{2\beta E_0[v^2]+v_t^2}} \quad ,
\end{equation}
we get
\begin{equation} \mylabel{e210}
f(x_{t+1}|v_{t})=\frac{2}{\sqrt{\pi}}\cdot
                 \frac{\Gamma(\beta +2)}{\Gamma(\beta +3/2)}
                 \left (1+x_{t+1}^2\right )^{-\beta -2}\quad .
\end{equation}
The conditional distribution of $x_{t+1}$ is independent of the previous
return $v_{t}$. From this we can determine $\beta$, since only at the
true $\beta$ the data for $x_{t+1}$ 
with different $v_{t}$ will collapse into a single scaling function 
$f(x_{t+1})$. This function can be compared with the predicted curve
\myref{e210}.\\
Knowing the two point function the autocorrelation of $v^q$  can be derived 
(see appendix \ref{A.3} ). The ratio
\begin{equation} \mylabel{e211}
R=\frac{E[v^q_{t+t_0}v^q_{t_0}]}{E_0^2[v^q]}=F(q/2,q/2;\ \beta +1;\ e^{-bt})
\end{equation}
involves again a hypergeometric function. For negative values of
$\alpha=q-\beta -1$ the variance of $v^q$ exists and we find the
autocorrelation function $C(t)$
\begin{equation} \mylabel{e212}
C_q(t)=\frac{E_0^2[v^q]}{\mbox{var}[v^q]}\cdot 
       \left (F(q/2,q/2;\ \beta +1;\ e^{-bt})-1\right  )\quad .
\end{equation}
If we expand for large $t$ the hypergeometric function we find with the leading
term an exponentially decay 
\begin{equation} \mylabel{e212a}
C_q(t)=\frac{q^2}{4(\beta +1)}\cdot e^{-bt}\quad .
\end{equation}
For small $b$ and finite $t$ the deviation from the exponential dependence
\myref{e212a} and the exact expression \myref{e212}
may be substantial. It is interesting to note that for $\alpha >0$ the
ratio $R$ becomes singular at $t=0$. Applying Eulers relation
to the hypergeometric function one observes for the ratio
\begin{equation} \mylabel{e213}
R=\frac{\Gamma(\beta +1)\Gamma(\alpha)}{\Gamma(q/2)^2}\cdot
   \left (bt\right )^{-\alpha}
\end{equation}
a power law in $t$. This does not mean a long memory since neither
the variance exists nor it consists in a large $t$ effect. The example
given by Lillo et al.\mycite{LMM} shows that Markov processes may lead
to power law in $t$ also with existing variance.

\section{Empirical Comparison with the DAX \label{Data}}
\begin{figure}[h]
\begin{center}
\epsfig{file=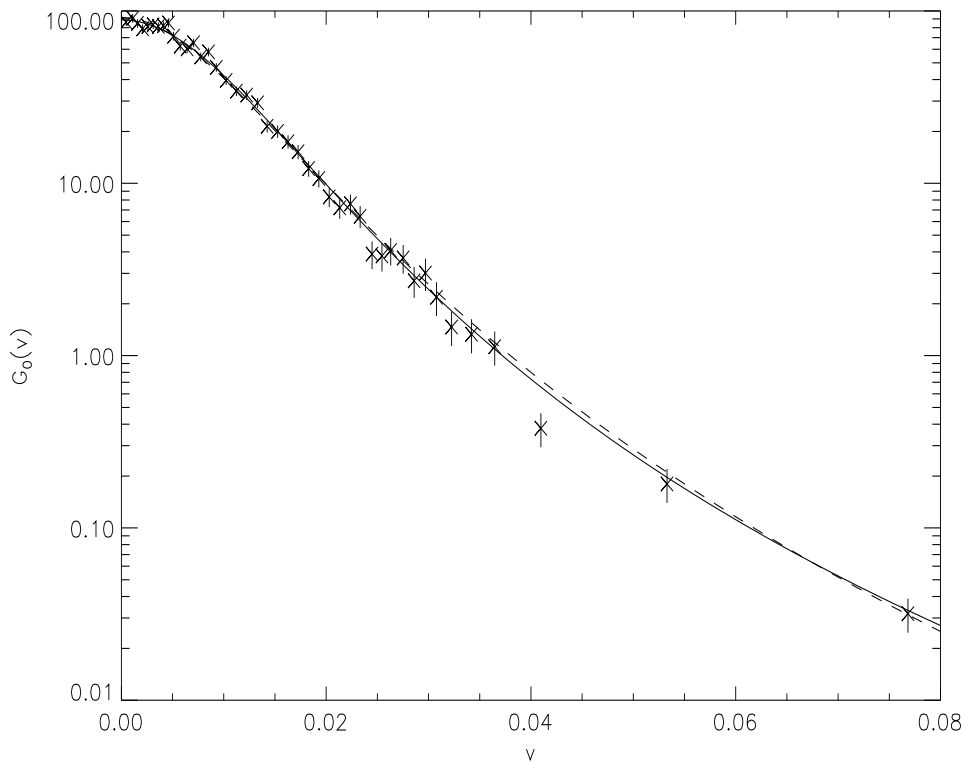,height=11cm,width=13cm,angle=0}
\caption{\label{fig1} \em The pdf from equ 
         \myref{e205} for equilibrium (solid line) 
         as function of the absolute return compared
         with the DAX from 1973-2002. The dashed line gives the result of
         the linear relation. }
\end{center}
\end{figure}

In this section we want to estimate the parameters of the model
described in the previous section from a time serie of absolute
returns of the DAX\mycite{DAX} during 1973-2002. The most efficient method
would be using a maximum likelihood fit. The bad analytical  properties
of $A$ make this task prohibitive unless $b$ is large. As less efficient
but tractable method to estimate $\beta$ and $r_0$ is to 
use $\chi^2$ fits to the unconditional pdf given in equ \myref{e205}.
In presence of correlations the sample of length $T$ can be still
considered as a representative subset of an infinite time serie if
$bT>>1$ holds. A small rate parameter $b$ may cause
a problem. To see this we divide the data in 5 subsamples of similar
statistics. Here and in all subsequent estimates we fix $r_0$ by
equ\myref{e207} in terms of $\beta$ and $E_0[v^2]$. We checked that in
each case a
free variation of $r_0$ did not alter the results. The estimates of $\beta$
in each sample are given in table \ref{table1}. They behave rather
erratic and $\chi^2$ probability is rather low indicating a poor
descripition of the data. This effect is due to the volatility clusters
which are by no means equally distributed over the subsamples.
\begin{table}
\begin{center}
\begin{tabular}[ht]{|c|cc|c|} \hline
Intervall & $\beta$ & $\Delta \beta$ & $\chi^2$/point \\ \hline
1/1973-10/1978 & 4.3   & 1.4   &  9/19 \\
11/1978-8/1984 & 3.3   & 1.0   & 31/18 \\
9/1984-7/1990  & 0.98  & 0.12  & 39/25 \\
8/1990-3/1996  & 1.04  & 0.15  & 41/22 \\
4/1996-2/2002  & 1.86  & 0.35  & 21/30 \\ \hline
1/1973-2/2002  & 1.115 & 0.068 & 44/43 \\ \hline
\end{tabular}
\caption{\label{table1} \em $\beta$ values and their statistical errors
         obtained by a fit in four different
         time intervalls. The last row gives the fit using all data. Column
         3 gives the $\chi^2$ value per data point. }
\end{center}
\end{table}
Taking all data one obtains a good $\chi^2$ value and
the following value of $\beta$
\begin{equation} \mylabel{e214}
\beta=1.115\; \pm \; 0.068\quad .
\end{equation}
In figure \ref{fig1} we compare the pdf from equ \myref{e205} with the 
empirical data.
The agreement is excellent with a $\chi^2$-probability of 60\%. The values
in the subsamples agree with \myref{e214} within two standard deviations.
The value \myref{e214} corresponds to a tail index of $n_T\sim 4$ which
is slightly larger than the the values obtained in other
financial markets \mycite{uni,IND}. As we discuss lateron this may be an effect
of not having reached the asymtotic regime in the data.\\
We can compare our model with a model discussed by Alfarano \mycite{Alf}
where the same herding mechanism
is used but the sqare root law replaced by the linear relation \myref{e122}.
The fit is  worse ( $\chi^2$-probability $\sim$ 1\% ), but in view of
possible systematic errors still acceptable. The difference stems mainly from 
the better description of low returns in our model. On a log scale the
fit in the linear model ( dashed line in figure \ref{fig1} ) can hardly
be distinguished from our model, although the
tail index  $n_T\sim 6$ is 50\% larger than in our case. 
A $\chi^2$- or maximum likelihood fit is dominated by the many
events with low $v$. To judge the 
description of extreme events we compare the predicted events with
$v>x\sqrt{E_0[v^2]}$ in both models with the observed number
of events. Since the largest return in figure \ref{fig1} correponds
to $x\sim 6$ the fits may not be sensible to extreme events with larger $x$.
The values for $x \ge 5$ are given in table \ref{table2}.
Despite of the large difference in the tail index both models agree with
the observation reasonably well with a preference for the 
square root law. Large 
differences occur only outside the observable region ($x>15$).
We also learn that there is no need for an extra mechanism for crash events
as proposed in \mycite{KAT}.\\
\begin{table}
\begin{center}
\begin{tabular}[ht]{|c|ccc|ccc|} \hline
&\multicolumn{3}{c|}{1985-2002}&\multicolumn{3}{c|}{1973-2002}\\ \hline
$ x$ & $N_{sqrt}$ & $N_{lin}$ & $N_{obs}$ &$N_{sqrt}$ & $N_{lin}$ & $N_{obs}$
 \\ \hline 
 5 & 8.1  & 7.6  & 11 & 13.8 & 13.0  & 24 \\
 6 & 3.9  & 3.4  &  6 &  6.6 &  5.7  &  9 \\
 7 & 2.1  & 1.6  &  3 &  3.6 &  2.8  &  6 \\
 9 & 0.8  & 0.5  &  1 &  1.3 &  0.8  &  1 \\
11 & 0.3  & 0.2  &  0 &  0.6 &  0.3  &  1 \\
15 & 0.09 & 0.03 &  0 &  0.2 &  0.06 &  0 \\ \hline
\end{tabular}
\caption{\em Predicted number of events with normalized absolute return
         larger than x (column 1) for the square root
         law (column 2 and 5) and the linear law (column 3 and 6) for
         two time intervalls. The observed number from DAX is given in
         column 4 and 7.  \label{table2}}
\end{center}
\end{table}
\begin{figure}[h]
\begin{center}
\epsfig{file=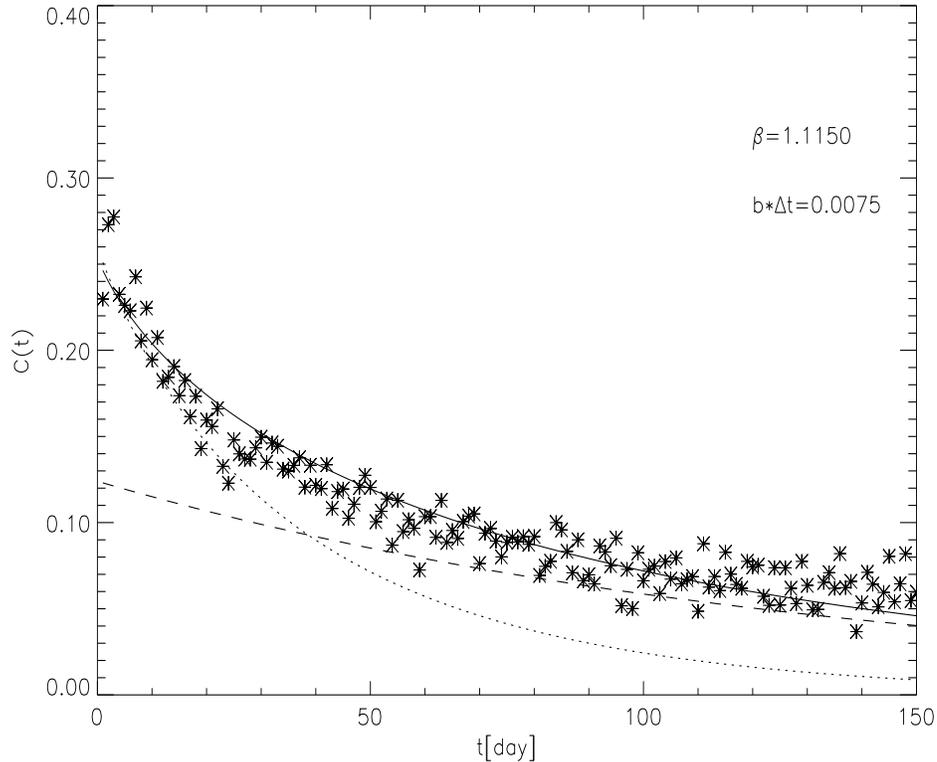,height=11cm,width=13cm,angle=0}
\caption{\label{fig2}\em  Correlation function for DAX. Solid line gives the
         prediction of equ \myref{e212}, the dashed line the asymtotic
         term \myref{e212a} and the dotted line the prediction of the 
         linear relation. }
\end{center}
\end{figure}
Both models can be distinguished if we consider the time dependence.
The parameter $b$ can be estimated from the common pdf for 
pairs $[v_{t+t_0},v_t]$ given in equ\myref{A16} of the appendix. We perform
maximum likelihood fits by adding the log likelihoods for all pairs with a
time lag $t_{0i}=[1,2,5,10,20]$. Maximizing $\sum_{i,t}\log\;f(v_{t+t_0i},v_t)$
with respect to $b$ and $\beta$ we obtain
\begin{equation} \mylabel{e215}
\beta=1.083\pm 0.020
\end{equation}
\begin{equation} \mylabel{e216}
b=0.0075\pm 0.0015\quad .
\end{equation}
The value of $\beta$ is compatibel with the estimate \myref{e214} from the
equlibrium pdf. The estimate \myref{e216} for $b$ implies a large decay time
$t_D=1/b\sim 133 [$day] in the order of 1/2 year. In the linear model only the
autocorrelation function can be computed. This model cannot describe the 
data over large range of time lags. Adjusting $b$ to describe the
autocorrelation for small time lags a value $b\sim 1/37 [\mbox{day}]_{-1}$
is obtained \mycite{Alf}. In figure \ref{fig2} we compare the autocorrelation 
function given in \myref{e212} (solid line) of our model with the values from
DAX. The agreement is good. The leading asymtotic term \myref{e212a} (dashed 
line) shows that the data cannot be described by an exponential behaviour.
The prediction of \mycite{Alf} (dotted line) should be valid only for small
time lags. Therefore the data for the autocorrelation prefer the
square root law.\\

\begin{figure}[h]
\begin{center}
\epsfig{file=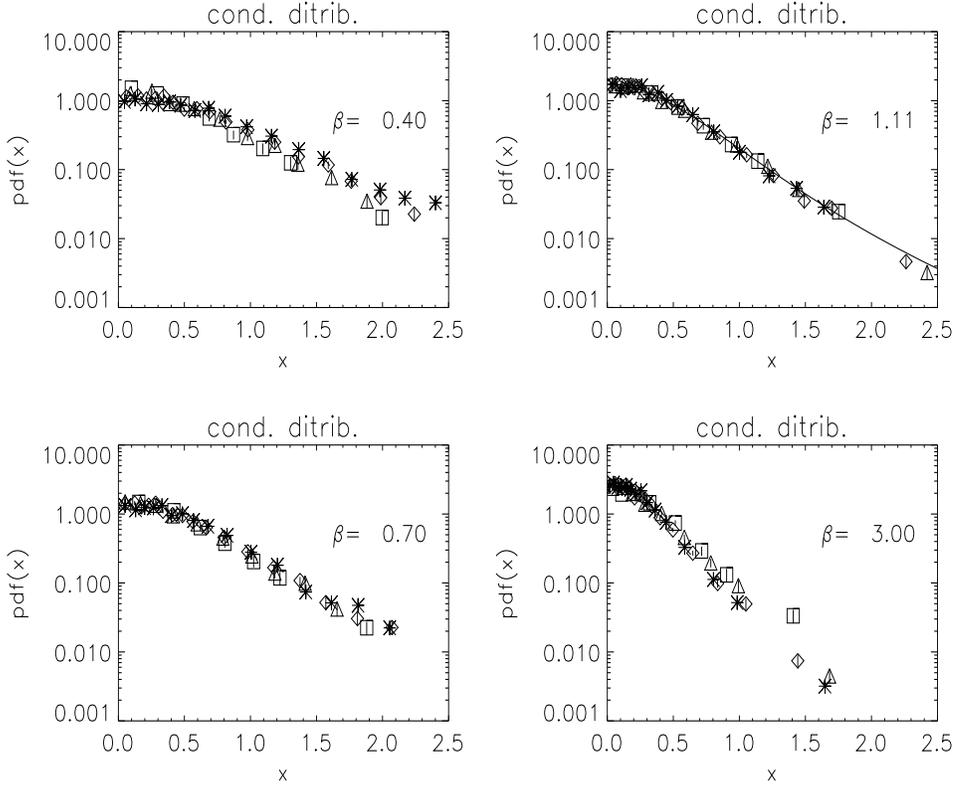,height=11cm,width=13cm,angle=0}
\caption{\label{fig3} \protect \em 
         Conditional pdf for the scaling variable $x_{t+1}$
         defined in (\ref{e217}) for various $\beta$ and $v_t$ intervalls.
         The solid line for $\beta=1.11$ gives the prediction \ref{e210}. 
         The intervalls of the normalized return $y$ from equ \ref{e217a}
         are denoted by stars (diamonds, triangles, squares) for \protect 
         $y\le$0.35 ( 0.35$\le y \le$ 0.9, 0.9$\le y\le$ 1.5, $y\ge$1.5 ).
       }
\end{center}
\end{figure}

Finally we test the scaling prediction, that the conditional pdf 
$f(v_{t+1}|v_t)$ is independent of $v_t$ if one uses the scaling variable
\begin{equation} \mylabel{e217}
x_{t+1}=\frac{v_{t+1}}{\sqrt{2\beta E_0[v^2]+v_t^2}}\quad .
\end{equation}
The assumption of small $b$ is very well satisfied with \myref{e216}.
In figure \ref{fig3} we show the distribution of $x$ obtained in four intervalls
of the normalized absolute return
\begin{equation} \mylabel{e217a}
y_t=\frac{v_t}{\sqrt{E_0[v^2]}}
\end{equation}
at $\beta=0.4,0.7,1.11,3.0$. For a $\beta \sim 1.1$ the data
collapse into a single curve. This curve
agrees with the predicted scaling function
\myref{e210} (solid line). From this agreement we conclude that the model
describes the correlation between $v_t$ and $v_{t+1}$ well, which is not
obvious from the autocorrelation due to the large fluctuation near $t=0$
seen in figure \ref{fig2}. A best fit to the scaling curve \myref{e210}
leads to $\beta=1.1 \pm 0.2$ which agrees with the previous estimations in
\myref{e214} and \myref{e215}. It is very remarkable that three very different
methods to determine $\beta$ lead to the same value.

\section{ Conclusions \label{concl}}
The herding model of Kirman has been considered in the limit 
that fundamentalists are much less affected by herding than 
noise traders. As a 
consequence the market consists of many noise traders and few fundamentalists
reflecting the conditions in a real market. Volatility cluster
occur if the fundamentalists dissapear from the market.
This is in contradiction to the results of
the Lux-Marchesi model, where fundamentalists dominate and volatility
clusters appear if chartists rise beyond 30\%. Augmenting the herding
model with Zhang's square root law between price changes and agent ratios
the model turns out to equivalent to the 3/2 model of statistical
volatility. This coincidence allows an interpretation of the parameters
in the latter model in terms of agent behaviour.
Due to the simplicity of our model 
various observables can be calculated analytically.
This is important in $\chi^2$ or maximum likelihood
estimations, where numerically accurate and computationally easy 
accessible expressions are required.\\
Comparing our model with empirical data we find good agreement with the
stylized facts derived from the DAX. The tail index of 4 is in harmony
with universality. Replacing the square root law by a linear relation
one gets a worse but still acceptable value of $\chi^2$, however the
tail index increases to 6. This shows that the index depends strongly
whether one uses a Pareto law in the squared return
$v^2$ (as in our model) or in $v$.
Universality of the tail index and the behaviour of extreme events favor
the square root law. The autocorrelation of the absolute return will 
decay exponentially at large time lags. The very small rate
constant $b$ ensures that this asymtotic behaviour is reached outside the
region of timelags where the data can be trusted. The reasonable 
agreement of the autocorrelation function of our model with the data
is much better than in the linear model. A quantitave measure of the
volatility cluster is expressed by a new scaling law for the pdf
conditional on the value of the previous day, which is again in good
agreement with the data.\\
The predictions based on a recursive solution of the combined Markov 
process of return and agent ratios will be presented in a future
publication.\\
{\bf Acknowlegments}: The author thanks Thomas Lux and Simone Alfarano
for stimulating discussions and valuable hints.

\begin{appendix}
\section{Asymmetric Kirman Model \label{A.1} }
In this section we give the changes of the formulae of \mycite{AWL}
in the case if one of the parameters describing the 
spontaneous change of oppinion becomes large. As pointed out in \mycite{AWL} 
the macroscopic time scale $\Delta t$ during the returns are observed 
does not need to coincide with the microscopic time scale $\Delta t_m$
over which the agents change oppinion. On the latter scale we have the FPE 
for $z=n_S/(n_S+n_F)$ derived in \mycite{AWL} from the transition 
probabilities \myref{e111}
\begin{equation} \mylabel{A1}
\frac{\Delta w(z,t)}{\Delta t_m}=\frac{\partial}{\partial z}
  \left [ za_2-(1-z)a_1+\frac{\partial}{\partial z}bz(1-z)\right ]w(z,t)\quad .
\end{equation}
For large $a_1$ z will be close to 1. Therefore we introduce the variable $u$ by
\begin{equation} \mylabel{A2}
u=\frac{a_1}{b}(1-z)\;=\; \frac{a_1}{b}\frac{n_F}{n_S+n_F}
\end{equation}
which remains finite. Transforming \myref{A1} into $u$ and neglecting terms
of the order $b/a_1$ we find
\begin{equation} \mylabel{A3}
\frac{\Delta w(u,t)}{\Delta t_m}=a_1\cdot D_u w(u,t)
\end{equation}
with the differential operator
\begin{equation} \mylabel{A4}
D_u=\frac{\partial}{\partial u}\left [u-\frac{a_2}{b}+
      \frac{\partial}{\partial u}u \right ]\quad .
\end{equation}
The real time  scale $\Delta t$ (f.e. days) is related to $\Delta t_m$ by
\begin{equation} \mylabel{A5}
\Delta t=\frac{a_1}{b}\cdot \Delta t_m\quad .
\end{equation}
The limit of large $a_1$ leads to small average values of the number of
fundamentalists. By the identification \myref{A5} the agent ratio can be
expressed in terms of the time scale
\begin{equation} \mylabel{A5a}
\frac{n_F}{n_S+n_F}=\frac{\Delta t_m}{\Delta t}\; u
\end{equation}
Inserting \myref{A5} into equ. \myref{A3} leads to
\begin{equation} \mylabel{A6}
\frac{\Delta w(u,t)}{\Delta t}=b\cdot D_u w(u,t)
\end{equation}
which is equivalent to equ. \myref{e116} replacing $\Delta w(u,t)/\Delta t$
by $\partial w(u,t)/\partial t$ and $a_2/b$ by $\beta +1$. The FPE \myref{e113}
( see \mycite{Ahn} ) can be solved by expanding $w(u,t)$ in terms of
Laguerre polynomials $L_n^\beta(u)$: 
\begin{equation} \mylabel{A7}
w(u,t)=\sum_{n=0}^{\infty} c_n(t)L_n^\beta(u)w_0(u)
\end{equation}
with the equilibrium pdf $w_0(u)$ from equ. \myref{e116}.
The functions $L_n^\beta(u)\cdot w_0(u)$ are right eigen functions of $D_u$
\begin{equation} \mylabel{A8}
D_u \cdot L_n^\beta(u)w_0(u)=-n\cdot L_n^\beta(u)w_0(u)\quad .
\end{equation}
The conditional pdf $A(u|\bar{u},z)$ follows from the generating function
of the Laguerre polynomials (see \mycite{Erd} ). The convolution property
\myref{e118} is proven by the orthogonality of $L_n^\beta$
\begin{equation} \mylabel{A9}
E_0[L_n^\beta(u)L_{n'}^\beta(u)]=\delta_{n,n'}
\end{equation}
with $E_0[.]$ the expectation value with respect to $w_0$.

\section{Models Outside the Critical Region \label{spin}  }
A general 
class of models can be characterized by the following cost functional
\begin{equation} \mylabel{C1}
S(\Phi)=-\frac{g+g_0}{N\sigma_0^2}\sum_{i<j}\Phi_i\Phi_j\; - \;
         \frac{\mu_0\Omega}{\sigma_0}\sum_{i}\Phi_i \quad .
\end{equation}
The minimum coupling $g_0$ ensures that the system is never in the disordered
state. $\sigma_0$ and $\mu_0$ set the scales for $\Phi$ and $\Omega$.
A suitable dynamic (Langevin equation, heat bath or Metropolis algorithm)
will bring the system into equilibrium, which is a Boltzmann distribution.
Expectation values are given by derivatives of the partition sum $Z$:
\begin{equation} \mylabel{C2}
Z=\prod_{i}\int d\mu(\Phi_i)\exp(-S(\Phi) \quad .
\end{equation}
For the measure $d\mu(\Phi)$ we can use either a spin or a compact model.
$\mu$ reads in these cases
\begin{equation} \mylabel{C3}
\int d\mu(\Phi)=\left\{ \begin{array}{l@{\mbox{\quad for\quad }}l}
\frac{1}{2}\sum_{\Phi=\pm \sigma_0} & \mbox{ for the Ising model} \\
\frac{1}{2\sigma_0}\int d\Phi  & \mbox{ for the compact model}
\end{array} \right .
\end{equation}
The expectation value of the return is given by
\begin{equation} \mylabel{C4}
v=E\left [\frac{1}{N}\sum_{i}\Phi_i\right ]=\frac{\sigma_0}{N\mu_0}
    \frac{\partial \ln Z}{\partial \Omega}
\end{equation}
To evaluate $Z$ we use the Gaus trick for large agent number $N$:
\begin{eqnarray} \mylabel{C5}
\exp(\frac{g+g_0}{N\sigma_0^2}\sum_{i<j}\Phi_i\Phi_j)&=&
 \sqrt{\frac{N}{2\pi (g+g_0)}}  \cdot \nonumber \\
 & &\int d\omega\exp\left (
               -\frac{N\omega^2}{2(g+g_0)}\;
               +\; \frac{\omega}{\sigma_0}\sum\Phi_i\right )
\end{eqnarray}
Inserting equ \myref{C5} into $Z$ the integration over $\Phi_i$ can
be carried out with the result
\begin{eqnarray} \mylabel{C6}
Z &=& \sqrt{\frac{N}{2\pi (g+g_0)}}\cdot  \nonumber \\
  & & \int d\omega\; \exp\left (
-N\left [\frac{\omega^2}{2(g+g_0)}\;-\;\ln C(\omega+\mu_0\Omega)\right ]
          \right ) \quad .
\end{eqnarray}
For the spin model we set $g_0=1$ and $C$ is given by
\begin{equation} \mylabel{C7}
C(x)=\cosh (x)\quad ,
\end{equation}
whereas in the compact model we use $g_0=3$ and the function $C$ as
\begin{equation} \mylabel{C8}
C(x)=\frac{1}{x}\cdot \tanh (x) \quad .
\end{equation}
In the large $N$-limit $Z$ is given by the maximum of the integrand at
$\omega_0$. For the spin model we get
\begin{equation} \mylabel{C9}
\omega_0=(g+1)\tanh (\omega_0+\mu_0\Omega) \quad .
\end{equation}
Replacing $\omega_0$ by the expectation value
$v=E[\Phi_i]=\omega_0\sigma_0/(g+1)$ we find the equation of state
\begin{equation} \mylabel{C10}
v=\sigma_0\tanh \left ((g+1)\frac{v}{\sigma_0}\; +\;\mu_0\Omega\right )
\end{equation}
To get the relation with the parameters of the critical model discussed
in section \ref{sec1a} we expand equ \myref{C10} around small $g$ and $\mu_0$
\begin{equation} \mylabel{C11}
\frac{1}{3\mu_0}\left (\frac{g+1}{\sigma_0}\right )^3\cdot v^3
\;-\;\frac{g}{\sigma_0\mu_0}\cdot v\;-\; \Omega=0
\end{equation}
Comparing coefficients with the equation of state \myref{e123b2} we get
two equations for the three parameters $g$, $\sigma_0$ and $\mu_0$. For
example the scale parameter $\mu_0$ for $\Omega$ can be chosen
arbitrarily. Measuring $\mu_0$ in units of $\Omega_0$ we set
\begin{equation} \mylabel{C12}
\mu_0=\frac{1}{(6g_0-3)\Omega_0}\cdot y^3
\end{equation}
and express $g$ and $\sigma_0$ in terms of the previous parameters
$\Omega_0$ and $\alpha_S$
\begin{equation} \mylabel{C13}
g=\frac{y^2}{3-y^2} \quad \mbox{and} \quad
  \sigma_0=\frac{\Omega_0}{\alpha_S}\frac{3}{y(3-y^2)}
\end{equation}
Inserting these parameters into equ \myref{C10} we obtain the equation
of state valid also outside the critical region
\begin{equation} \mylabel{C14}
\frac{\alpha_S}{\Omega_0}\cdot v=\frac{3}{(3-y^2)y}\cdot
        \tanh \left (\xi\right )
\end{equation}
with
\begin{equation} \mylabel{C15}
\xi=\frac{\alpha_S}{\Omega_0}\cdot v\; +\; \frac{y^3}{6g_0-3}
    \cdot \frac{\Omega}{\Omega_0}
\end{equation}
The parameter $y$ with $0<y<\sqrt{3}$ describes the distance from the critical
behaviour obtained for $y\to 0$. The saturation of the return as function of
$\Omega$ determines the value of $y$. The exact form is model dependent.
The analogous calculation for the compact model leads instead of \myref{C14} to
\begin{equation} \mylabel{C16}
\frac{\alpha_S}{\Omega_0}\cdot v=\frac{45}{(15-y^2)y}\cdot
        \left (\coth (\xi )\;-\; \frac{1}{\xi} \right )
\end{equation}
It has the same behaviour for $y\to 0$, but extrapolates hyperbolically
for small $1/\Omega$ to a constant instead of the exponential
behaviour obtained from \myref{C14}.

\section{Two Point Function of the Probability Density \label{A.2}  }
From the expansion of the Bessel function in \myref{e117} we
obtain the following representation for $A$ with $z=\exp(-bt)$
\begin{eqnarray} \mylabel{A12}
A(u|u',z)&=&\frac{u^\beta}{(1-z)^{\beta +1}}\; 
          \exp\left (-\frac{u}{1-z}-\frac{zu'}{1-z} \right ) \cdot\nonumber \\
         &\; & \sum_{m=0}^{\infty} \left (\frac{uu'z}{(1-z)^2} \right )^m
          \; \frac{1}{m!}\; \frac{1}{(\beta +m)!}\quad .
\end{eqnarray}
Inserting \myref{A12} and the noise distribution \myref{e202} into 
equ\myref{e208} we obtain after integration over $u$ and $u'$
\begin{eqnarray} \mylabel{A13}
f(v_{t+t_0},v_{t_0})&=&\frac{N_0^2}{\beta !}\;
       (1-z)^{\beta+2}\; \sum_{m=0}^{\infty}
       \frac{\left ((\beta+m+1/2)!\right )^2}{m!(\beta+m)!}\; 
       z^m \cdot\nonumber \\
       & & \left [(1+(1-z)\frac{v_{t+t_0}^2}{2r_0^2})\cdot
              (1+(1-z)\frac{v_{t_0}^2}{2r_0^2}) \right ]^{-(\beta+m+3/2)}\quad .
\end{eqnarray}
The sum corresponds to the serie expansion of the hypergeometric function
\begin{equation} \mylabel{A14}
F(a,b;c;x)=\sum_{m=0}^{\infty}\frac{(a+m-1)!}{(a-1)!}\;\frac{(b+m-1)!}{(b-1)!}\;
           \frac{(c-1)!}{(c+m-1)!}\; \frac{x^m}{m!}\quad .
\end{equation}
At special arguments $F(a,b;c;0)=1$ and 
$F(a,b;c+1;1)=c!(c-a-b)!/((c-a)!(c-b)!)$ holds. Using the scaling variable
\begin{equation} \mylabel{A15}
\zeta=\frac{1}{2r_0^2} \left (
      v_{t+t_0}^2+v_{t_0}^2+(1-z)\; \frac{v_{t+t_0}^2v_{t_0}^2}{2r_0^2}
      \right )
\end{equation}
we get
\begin{eqnarray} \mylabel{A16}
f(v_{t+t_0},v_{t_0})&=&\left (\frac{N_0(\beta+1/2)!}{\beta !}\right )^2
       \frac{(1-z)^{\beta+2}}{(1+(1-z)\zeta)^{\beta+3/2}}\cdot\nonumber \\
      &\;& F\left (\beta+3/2,\beta+3/2;\;\beta+1;\;
                  \frac{z}{(1+(1-z)\zeta)}\right )\quad .
\end{eqnarray}
With Euler's relation $F(a,b;c;x)=(1-x)^{c-a-b}F(c-a,c-b;\; c;\; x)$ 
equ\myref{A16} can be rewritten as
\begin{eqnarray} \mylabel{A17}
f(v_{t+t_0},v_{t_0})&=&\left (\frac{N_0(\beta+1/2)!}{\beta !}\right )^2
       \frac{\sqrt{1+(1-z)\zeta}}{(1+\zeta)^{\beta+2}}\cdot\nonumber \\
       &\;&F\left (-1/2,-1/2;\;\beta+1;\;\frac{z}{(1+(1-z)\zeta)}\right )\quad .
\end{eqnarray}
If terms of order $1-z$ are negligeable and $t=1$ 
$\zeta$ agrees with the scaling 
variable $x^2_{t+1}$ from equ \myref{e209a}. Setting in $F$ the argument to 1
leads to
\begin{equation} \mylabel{A18}
f(v_{t+1},v_{t})=\frac{2(\beta+1)}{\pi r_0^2} \;
                 \left (1+x_{t+1}^2\right )^{-\beta-2}\quad .
\end{equation}
This equation is equivalent to \myref{e209} if we use
\begin{equation} \mylabel{A19}
f(x_{t+1}|v_{t})=\frac{dv_{t+1}}{dx_{t+1}}\; 
                 \frac{f(v_{t+1},v_{t})}{G_0(v_{t})}
\end{equation}
with the equilibrium pdf \myref{e205} for $v_{t}$.

\section{Autocorrelation \label{A.3} }
The easiest way to derive $E[v_{t+t_0}^q\cdot v_{t_0}^q] $ is to perform
in equ\myref{e208} first the $v$ integrations using 
\begin{equation} \mylabel{A20}
\int_0^\infty g(v|u)\; v^q\; dv=E_0[v^q]\frac{\beta!}{(\beta-q/2)!}\;u^{-q/2}
\end{equation}
with the equilibrium moments
\begin{equation} \mylabel{A21}                             
E_0[v^q]=\sqrt{\frac{2^q}{\pi}}\; r_0^q\cdot 
         \frac{(\beta-q/2)!((q-1)/2)!}{\beta!} \quad .
\end{equation}
Inserting as in the previous section the expansion \myref{A12} and performing
the integrations over $u$ and $u'$ we obtain for the ratio
\begin{eqnarray} \mylabel{A22}                              
R &=& \frac{E[v_{t+t_0}^q\; v_{t_0}^q]}{E_0[v^q]^2} \nonumber \\
  &=& (1-e^{-bt})^{\beta+1-q}\cdot\sum_{m=0}^{\infty}
      \left (\frac{(\beta+m-q/2)!}{(\beta-q/2)!}\right )^2\cdot
      \frac{\beta!\cdot e^{-btm}}{m!(\beta+m)!}
\end{eqnarray}
Again the sum is a hypergeometric function leading to
\begin{equation} \mylabel{A23}                              
R=(1-e^{-bt})^{\beta+1-q}\cdot 
   F\left (\beta+1-q/2,\beta+1-q/2;\;\beta+1;\; e^{-bt} \right )
\end{equation}
With the help of Euler's relation we get finally
\begin{equation} \mylabel{A24}                              
R=F\left (q/2,q/2;\beta+1;e^{-bt} \right )\quad .
\end{equation}
For large $bt$ expansion of $F$ up to linear terms leads to
\begin{equation} \mylabel{A25}                              
R=1+\frac{q^2}{4(\beta +1)}\cdot e^{-bt}\quad .
\end{equation}
For $bt\to 0$ and $\beta+1-q<0$ 
the expectation value $R$ becomes singular. Replacing in
equ\myref{A23} the function $F$ by its value at argument 1 we find the
power law given in equ \myref{e213}

\end{appendix}

\end{document}